\documentclass[twocolumn,showpacs,preprintnumbers]{revtex4}
\usepackage{graphicx}
\usepackage{dcolumn}
\usepackage{bm}

\begin{document}

\preprint{}
\title{Realization of universal quantum cloning with SQUID qubits in a cavity%
}
\author{Jian Yang}
\author{Ya-Fei Yu}
\author{Zhi-Ming Zhang}
\thanks{Corresponding author. Electronic address: zmzhang@scnu.edu.cn}
\affiliation{Laboratory of Photonic Information Technology, School of Information and
Photonelectronic Science and Engineering, South China Normal University,
Guangzhou 510006, China. }
\date{\today }

\begin{abstract}
We propose a scheme to realize $1\rightarrow 2$ universal quantum cloning
machine (UQCM) with superconducting quantum interference device (SQUID)
qubits, embeded in a high-Q cavity. CNOT operations are derived to present
our scheme, and the two-photon Raman resonance processes are used to
increase the operation rate. Compared with previous works, our scheme has
advantages in the experimental realization and further utilization.
\end{abstract}

\maketitle

\section{Introduction}

The field of quantum computation and quantum information has grown so
rapidly in the recent years \cite{Nielsen2001}. Quantum mechanics provides
us powerful tools to solve specific problems, while it also imposes extra
limitations. Duplication is one of the most common operations on the
classical computer, however, the so-called \textit{quantum no cloning theorem%
} tells us that one can not copy arbitrary unknown quantum states perfectly
\cite{Wootters1982}. However, several schemes have been proposed to achieve
realistic quantum cloning machines, which generate copies of arbitrary input
qubits approximately, or probabilistically [3-8].

The universal quantum cloning machine (UQCM) is first discussed by Buz\u{e}k
and Hillery \cite{Buzek1996}, which provides identical fidelities to
arbitrary input qubits. The progress on quantum cloning is reviewed by
Scarani et al. \cite{Scarani2005}. The optimal $1\rightarrow 2$ UQCM
performs the transformations:

\begin{eqnarray}
\left\vert +\right\rangle \left\vert \Sigma \right\rangle &\rightarrow &%
\sqrt{\frac{2}{3}}\left\vert +\right\rangle \left\vert +\right\rangle
\left\vert A_{\bot }\right\rangle +\sqrt{\frac{1}{3}}\left\vert \Phi
\right\rangle \left\vert A\right\rangle , \\
\left\vert -\right\rangle \left\vert \Sigma \right\rangle &\rightarrow &%
\sqrt{\frac{2}{3}}\left\vert -\right\rangle \left\vert -\right\rangle
\left\vert A\right\rangle +\sqrt{\frac{1}{3}}\left\vert \Phi \right\rangle
\left\vert A_{\bot }\right\rangle ,  \nonumber
\end{eqnarray}%
where qubits are encoded in the basis $\{\left\vert \pm \right\rangle \}$,
while $\left\vert \Sigma \right\rangle $ is the initial state of blank
copies and ancilla qubits, $\left\vert A_{\bot }\right\rangle $ and $%
\left\vert A\right\rangle $ are the final ancilla states, and $\left\vert
\Phi \right\rangle =\left( \left\vert +\right\rangle \left\vert
-\right\rangle +\left\vert -\right\rangle \left\vert +\right\rangle \right) /%
\sqrt{2}$. After this operation, each copies has a fidelity of 5/6 \cite%
{Buzek1996}, when compared to the input state. Based on cavity QED, Milman
et al. \cite{Milman2003} and Zou et al. \cite{Zou3003} have proposed
different schemes to realize UQCM. However, at least two cavities or two
cavity-modes are needed in these works. Inspired by paper \cite{Yang2003},
we turn to construct UQCM with SQUID embed in a high-Q cavity, based on
which, many quantum information processing schemes have been proposed
[13-15]. Our motivation is as follows:

\textit{(1)} Unlike Rydberg atom used in cavity-QED, which is a "flying"
qubit, SQUID embed in a high-Q cavity is "static", so it is comparatively
easy to adjust its coupling coefficient to the cavity field, or to
manipulate it with classical field or quantized field. In our scheme, only
one cavity is needed, which accords with the current experimental conditions.

\textit{(2)} The level spacing of SQUID can be adjusted by changing the
external flux or the critical current \cite{Yang2003}, so we can easily
"turn-on" or "turn-off" the interaction between the SQUID and the cavity
field. Atom-velocity-selected device and passing qubits detection are not
needed in our scheme.

This paper is organized as follows. In Sec. II, we review the SQUID driven
by quantized or classical microwave fields. CNOT gate and two specific
processes are also constructed. In Sec. III, we discuss the detail of
realizing the UQCM. A brief discussion and conclusions are presented in Sec.
IV.

\section{Manipulation of SQUID}

We assume that the SQUID, considered throughout this paper, has a $\Lambda $%
-type three-level structure, as shown in FIG.1. The SQUID qubit is embed in
a high-Q cavity and can be manipulated by the cavity field as well as the
microwave pulses. In this section, we review the effective Hamiltonian of
the SQUID driven by a quantized or classical field. More details of SQUID
and Josephson-junction are discussed by Makhlin et al. \cite{Makhlin2001}.
Specific processes as well as CNOT gate are also derived to realize the UQCM.

\subsection{Resonant interaction between SQUID and a cavity field.}

Consider a $\Lambda $-type three-level SQUID embeded in a high-Q cavity. If
its $\left\vert g\right\rangle \leftrightarrow \left\vert e\right\rangle $
transition is resonant with the cavity field, the Hamiltonian in the
interaction picture under rotating-wave approximation can be written as \cite%
{Yang2003}:%
\begin{equation}
H_{I}=\lambda \left[ a^{\dag }\left\vert g\right\rangle \left\langle
e\right\vert +a\left\vert e\right\rangle \left\langle g\right\vert \right] .
\end{equation}%
where $\lambda $ is the effective coupling constant \cite{Yang2003}, and $%
a^{\dag }$ and $a$ are the creation and annihilation operator of the cavity
field, respectively. The evolution of this system can be easily derived as:%
\begin{eqnarray}
\left\vert g,1\right\rangle &\rightarrow &\cos (\lambda t)\left\vert
g,1\right\rangle -i\sin (\lambda t)\left\vert e,0\right\rangle , \\
\left\vert e,0\right\rangle &\rightarrow &\cos (\lambda t)\left\vert
e,0\right\rangle -i\sin (\lambda t)\left\vert g,1\right\rangle .  \nonumber
\end{eqnarray}

\subsection{Resonant interaction between SQUID and a classical microwave
pulse}

\bigskip Consider a $\Lambda $-type three-level SQUID driven by a classical
microwave pulse and suppose its $\left\vert g\right\rangle \leftrightarrow
\left\vert e\right\rangle $ transition is resonant with the classical field.
Then the effective Hamiltonian in the interaction picture under
rotating-wave approximation can be written as \cite{Yang2003}:
\begin{equation}
H_{I}=\Omega _{ge}\left( \left\vert g\right\rangle \left\langle e\right\vert
+\left\vert e\right\rangle \left\langle g\right\vert \right) ,
\end{equation}%
where $\Omega _{ge}$ is the effective coupling constant. The evolution of
this system can be written as \cite{Yang2003}:%
\begin{eqnarray}
\left\vert g\right\rangle &\rightarrow &\cos (\Omega _{ge}t)\left\vert
g\right\rangle -i\sin (\Omega _{ge}t)\left\vert e\right\rangle , \\
\left\vert e\right\rangle &\rightarrow &\cos (\Omega _{ge}t)\left\vert
e\right\rangle -i\sin (\Omega _{ge}t)\left\vert g\right\rangle .  \nonumber
\end{eqnarray}%
Similarly, if the $\left\vert i\right\rangle \leftrightarrow \left\vert
e\right\rangle $ transition is resonant with the microwave pulse, and the
other transitions are far-off resonant, the evolution of SQUID can be
written as \cite{Yang2003}:%
\begin{eqnarray}
\left\vert i\right\rangle &\rightarrow &\cos (\Omega _{ie}t)\left\vert
i\right\rangle -i\sin (\Omega _{ie}t)\left\vert e\right\rangle , \\
\left\vert e\right\rangle &\rightarrow &\cos (\Omega _{ie}t)\left\vert
e\right\rangle -i\sin (\Omega _{ie}t)\left\vert i\right\rangle .  \nonumber
\end{eqnarray}%
where $\Omega _{ie}$ is the effective coupling constant.

\subsection{Interaction between SQUID and two microwave pulses with large
detuning}

Consider a $\Lambda $-type three-level SQUID driven by two classical
microwave pulses 1 and 2 (depicted in FIG. 1). This system constructs a
two-photon Raman resonance, if the $\left\vert g\right\rangle
\leftrightarrow \left\vert e\right\rangle $ transition and the $\left\vert
i\right\rangle \leftrightarrow \left\vert e\right\rangle $ transition are
coupled to the microwave pulses 1 and 2 with identical detunings, i.e. $%
\Delta =\omega _{ge}-\omega _{1}=\omega _{ie}-\omega _{2}$, where $\omega
_{1}$ and $\omega _{2}$ are the frequencies of two microwave pulses. In the
case of large detuning, the upper state $\left\vert e\right\rangle $ can be
eliminated adiabatically, and the evolution of this system follows \cite%
{Yang2004}:%
\begin{eqnarray}
\left\vert g\right\rangle &\rightarrow &\cos \left( \lambda ^{\prime
}t\right) \left\vert g\right\rangle \\
&&+e^{-i\omega _{gi}t}e^{-i\left( \phi _{1}-\phi _{2}-\pi /2\right) }\sin
(\lambda ^{\prime }t)\left\vert i\right\rangle ,  \nonumber \\
\left\vert i\right\rangle &\rightarrow &e^{i\left( \phi _{1}-\phi _{2}+\pi
/2\right) }\sin \left( \lambda ^{\prime }t\right) \left\vert g\right\rangle
\nonumber \\
&&+e^{-i\omega _{gi}t}\cos (\lambda ^{\prime }t)\left\vert i\right\rangle ,
\nonumber
\end{eqnarray}%
where $\phi _{1}$ and $\phi _{2}$ are the initial phases of two classical
microwave pluses, respectively, and $\lambda ^{\prime }$ is the effective
coupling constant.

Now we turn to construct two processes based on Eq. (7), which are further
needed in our universal cloning scheme.

\textbf{Process 1}: In this process, we want to realize the transformation:%
\begin{eqnarray}
\left\vert +\right\rangle &\rightarrow &-\left\vert i\right\rangle , \\
\left\vert -\right\rangle &\rightarrow &\left\vert g\right\rangle ,
\nonumber
\end{eqnarray}%
where $\left\vert \pm \right\rangle =\left( \left\vert i\right\rangle \pm
\left\vert g\right\rangle \right) /\sqrt{2},$ and this can be realized with
two steps:

\textit{Step (1)}: Apply two microwave pulses to SQUID with $\phi _{1}-\phi
_{2}=3\pi /2$, and adjust the interaction time to be $t_{1}=3\pi /4\lambda
^{\prime }$.

\textit{Step (2)}: Turn off the classical pulse, and let the SQUID undergo a
free evolution for a time $t_{2}$, such that $\omega _{gi}\left(
t_{1}+t_{2}\right) =2m\pi $ ($m$ is an integer).

This process can be detailed as:%
\begin{eqnarray}
\left\vert g\right\rangle &\rightarrow &\frac{-1}{\sqrt{2}}\left\vert
g\right\rangle -\frac{e^{-i\omega _{gi}t_{1}}}{\sqrt{2}}\left\vert
i\right\rangle \\
&\rightarrow &\frac{-1}{\sqrt{2}}\left( \left\vert i\right\rangle
+\left\vert g\right\rangle \right) ,  \nonumber \\
\left\vert i\right\rangle &\rightarrow &\frac{1}{\sqrt{2}}\left\vert
g\right\rangle -\frac{e^{-i\omega _{gi}t_{1}}}{\sqrt{2}}\left\vert
i\right\rangle  \nonumber \\
&\rightarrow &\frac{-1}{\sqrt{2}}\left( \left\vert i\right\rangle
-\left\vert g\right\rangle \right) .  \nonumber
\end{eqnarray}

\textbf{Process 2: }In this process, we want to construct a transformation,
which has the form:%
\begin{eqnarray}
\left\vert g\right\rangle &\rightarrow &\left\vert -\right\rangle =\frac{1}{%
\sqrt{2}}\left( \left\vert i\right\rangle -\left\vert g\right\rangle \right)
, \\
\left\vert i\right\rangle &\rightarrow &-\left\vert +\right\rangle =\frac{-1%
}{\sqrt{2}}\left( \left\vert i\right\rangle +\left\vert g\right\rangle
\right) .  \nonumber
\end{eqnarray}%
and this can be achieved by two steps:

\textit{Step (1)}: Apply two microwave pulses to SQUID with $\phi _{1}-\phi
_{2}=\pi /2$, and adjust the interaction time to be $t_{1}^{\prime }=3\pi
/4\lambda ^{\prime }$;

\textit{Step (2)}: Turn off the classical pulse, and let the SQUID undergo a
free evolution for a time $t_{2}^{\prime }$, such that $\omega _{gi}\left(
t_{1}^{\prime }+t_{2}^{\prime }\right) =2m\pi $ ($m$ is an integer).

This process can be summarized as:%
\begin{eqnarray}
\left\vert g\right\rangle &\rightarrow &\frac{-1}{\sqrt{2}}\left\vert
g\right\rangle +\frac{e^{-i\omega _{gi}t_{1}^{\prime }}}{\sqrt{2}}\left\vert
i\right\rangle \rightarrow \left\vert -\right\rangle , \\
\left\vert i\right\rangle &\rightarrow &\frac{-1}{\sqrt{2}}\left\vert
g\right\rangle -\frac{e^{-i\omega _{gi}t_{1}^{\prime }}}{\sqrt{2}}\left\vert
i\right\rangle \rightarrow -\left\vert +\right\rangle ,  \nonumber
\end{eqnarray}

We can adjust the evolution time $t_{2}$ and $t_{2}^{\prime }$ to satisfy $%
t_{1}+t_{2}=t_{1}^{\prime }+t_{2}^{\prime }=2m\pi $. In this case, the two
processes cost the same time.

\subsection{CNOT Gate}

CNOT is one of the most essential operations in quantum information, and it
also plays important role in our scheme. Inspired by previous work \cite%
{Raimond2001}, which presents CNOT gate in the cavity-QED with Rydberg atom,
here we propose a scheme to achieve CNOT operation based on the interaction
between SQUID and the cavity field, as discussed in Section II A. Suppose
that the information is coded on the $\left\vert g\right\rangle $ and $%
\left\vert i\right\rangle $ levels of the SQUID, which is embed in a high-Q
cavity, and its $\left\vert g\right\rangle \leftrightarrow \left\vert
e\right\rangle $ transition is set to be resonant with the cavity field. The
SQUID is initially in an arbitrary state $\left\vert \psi \right\rangle
=\alpha _{s}\left\vert +\right\rangle +\beta _{s}\left\vert -\right\rangle $%
, and the cavity is prepared in a superposition $\left\vert \varphi
\right\rangle =\alpha _{f}\left\vert 0\right\rangle +\beta _{f}\left\vert
1\right\rangle $, where $\left\vert \pm \right\rangle _{s}=\left( \left\vert
i\right\rangle _{s}\pm \left\vert g\right\rangle _{s}\right) /\sqrt{2}$ ,and
$\left\vert \alpha _{s}\right\vert ^{2}+\left\vert \beta _{s}\right\vert
^{2}=\left\vert \alpha _{f}\right\vert ^{2}+\left\vert \beta _{f}\right\vert
^{2}=1$. The evolution of this system follows Eq. (3), and after a time $%
t=\pi /\lambda $, we have:%
\begin{eqnarray}
\left\vert \pm \right\rangle _{s}\left\vert 0\right\rangle _{f} &\rightarrow
&\left\vert \pm \right\rangle _{s}\left\vert 0\right\rangle _{f}, \\
\left\vert \pm \right\rangle _{s}\left\vert 1\right\rangle _{f} &\rightarrow
&\left\vert \mp \right\rangle _{s}\left\vert 1\right\rangle _{f},  \nonumber
\end{eqnarray}%
which corresponds to a CNOT operation. In this scheme, the cavity field is
the control qubit, and the SQUID is the target qubit.

\section{Universal quantum cloning machine}

In this section, we turn to describe the detail of the UQCM. Suppose that
three identical SQUIDs are embeded in a high-Q cavity (depicted in FIG. 2).
To make the couplings between the SQUIDs and the cavity field are the same,
we need to adjust the positions of every SQUID such that the magnetic
components $\mathbf{B}(\mathbf{r}_{1},t)$, $\mathbf{B}(\mathbf{r}_{2},t)$,
and $\mathbf{B}(\mathbf{r}_{3},t)$ of the cavity field, imposed on SQUIDs,
are the same. More technical considerations have been discussed in the
previous papers \cite{Yang2003}\cite{Zhang2006}. The cavity is initially
prepared in the vacuum state $\left\vert 0\right\rangle _{f}$, and SQUIDs
are in the state $\left\vert g\right\rangle _{1}\left\vert g\right\rangle
_{2}\left\vert g\right\rangle _{3}$.

Let SQUID1 be prepared in an arbitrary state $\left\vert \psi \right\rangle
_{1}=\alpha \left\vert +\right\rangle _{1}+\beta \left\vert -\right\rangle $
with two classical microwave pulses, where $\left\vert \alpha \right\vert
^{2}+\left\vert \beta \right\vert ^{2}=1$. The UQCM is to achieve the
transformation described by Eq. (1).

\textit{Step (1)}: Let\textit{\ }SQUID2 be driven by a classical microwave
pulse. Following Eq. (5), we have:%
\begin{equation}
\left\vert \varphi \right\rangle _{2}=\sqrt{\frac{2}{3}}\left\vert
g\right\rangle _{2}+i\sqrt{\frac{1}{3}}\left\vert e\right\rangle _{2}.
\end{equation}%
\

\textit{Step (2)}: Adjust the level spacing of SQUID2 to make its $%
\left\vert g\right\rangle \leftrightarrow \left\vert e\right\rangle $
transition resonant with the cavity field. Undergoing a $t=\pi /2\lambda $
evolution, the state of SQUID2\ and the cavity field is changed to:

\begin{equation}
\left\vert \varphi \right\rangle _{2}\left\vert 0\right\rangle
_{f}\rightarrow \left\vert g\right\rangle _{2}\left( \sqrt{\frac{2}{3}}%
\left\vert 0\right\rangle _{f}+\sqrt{\frac{1}{3}}\left\vert 1\right\rangle
_{f}\right) .
\end{equation}

\textit{Step (3)}: Turn-off SQUID2 (i.e. making it does not interact with
the field) , and let the $\left\vert g\right\rangle \leftrightarrow
\left\vert e\right\rangle $ transition of SQUID1 be coupled to the cavity
field. With a CNOT operation detailed in Eq. (12), the system is changed to:
\begin{eqnarray}
&&\sqrt{\frac{2}{3}}\left( \alpha \left\vert +\right\rangle _{1}+\beta
\left\vert -\right\rangle _{1}\right) \left\vert g\right\rangle
_{2}\left\vert 0\right\rangle _{f}+ \\
&&\sqrt{\frac{1}{3}}\left( \alpha \left\vert -\right\rangle _{1}+\beta
\left\vert +\right\rangle _{1}\right) \left\vert g\right\rangle
_{2}\left\vert 1\right\rangle _{f}.  \nonumber
\end{eqnarray}

\textit{Step (4)}: Turn-off SQUID1, and let the SQUID2 interact with the
cavity field again. After a time $t=\pi /4\lambda $, the system is changed
to
\begin{eqnarray}
&&\sqrt{\frac{2}{3}}\left( \alpha \left\vert +\right\rangle _{1}+\beta
\left\vert -\right\rangle _{1}\right) \left\vert g\right\rangle
_{2}\left\vert 0\right\rangle _{f}+ \\
&&\sqrt{\frac{1}{6}}\left( \alpha \left\vert -\right\rangle _{1}+\beta
\left\vert +\right\rangle _{1}\right) \cdot \left( \left\vert g\right\rangle
_{2}\left\vert 1\right\rangle _{f}-i\left\vert e\right\rangle _{2}\left\vert
0\right\rangle _{f}\right) .  \nonumber
\end{eqnarray}

\textit{Step (5)}: Turn-off SQUID2, and let SQUID3 interact with the cavity
field for a time $t=\pi /2\lambda $. We get:
\begin{eqnarray}
&&[\sqrt{\frac{2}{3}}\left( \alpha \left\vert +\right\rangle _{1}+\beta
\left\vert -\right\rangle _{1}\right) \left\vert g\right\rangle
_{2}\left\vert g\right\rangle _{3}- \\
&&i\sqrt{\frac{1}{6}}\left( \alpha \left\vert -\right\rangle _{1}+\beta
\left\vert +\right\rangle _{1}\right) \cdot \left( \left\vert g\right\rangle
_{2}\left\vert e\right\rangle _{3}+\left\vert e\right\rangle _{2}\left\vert
g\right\rangle _{3}\right) ]\left\vert 0\right\rangle _{f}.  \nonumber
\end{eqnarray}

\textit{Step (6)}: Let SQUID2 and SQUID3 be driven with identical classical
microwave pulses on its $\left\vert g\right\rangle \leftrightarrow
\left\vert i\right\rangle $ transition, respectively. After a time $t=\pi
/2\Omega _{ie}$, we get:%
\begin{eqnarray}
&&\sqrt{\frac{2}{3}}\left( \alpha \left\vert +\right\rangle _{1}+\beta
\left\vert -\right\rangle _{1}\right) \left\vert g\right\rangle
_{2}\left\vert g\right\rangle _{3}- \\
&&\sqrt{\frac{1}{6}}\left( \alpha \left\vert -\right\rangle _{1}+\beta
\left\vert +\right\rangle _{1}\right) \left( \left\vert g\right\rangle
_{2}\left\vert i\right\rangle _{3}+\left\vert i\right\rangle _{2}\left\vert
g\right\rangle _{3}\right) .  \nonumber
\end{eqnarray}

\textit{Step (7)}: Then we impose Process 1 (described in section II C) on
SQUID1, and Process 2 on SQUID2 and SQUID3, simultaneously. The system will
be transformed to:%
\begin{eqnarray}
&&\sqrt{\frac{2}{3}}\left( -\alpha \left\vert i\right\rangle _{1}+\beta
\left\vert g\right\rangle _{1}\right) \left\vert -\right\rangle
_{2}\left\vert -\right\rangle _{3}+ \\
&&\sqrt{\frac{1}{3}}\left( \alpha \left\vert g\right\rangle _{1}-\beta
\left\vert i\right\rangle _{1}\right) \left\vert \Phi \right\rangle _{23},
\nonumber
\end{eqnarray}%
where $\left\vert \Phi \right\rangle _{23}=\left( \left\vert +\right\rangle
_{2}\left\vert -\right\rangle _{3}+\left\vert -\right\rangle _{2}\left\vert
+\right\rangle _{3}\right) /\sqrt{2}$.

\textit{Step (8)}: Let SQUID1 undergo a $\left\vert i\right\rangle
_{1}\rightarrow -i\left\vert e\right\rangle _{1}$ transformation with a
classical microwave pulse. Following Eq. (13), we have:
\begin{eqnarray}
&&\sqrt{\frac{2}{3}}\left( \alpha i\left\vert e\right\rangle _{1}+\beta
\left\vert g\right\rangle _{1}\right) \left\vert -\right\rangle
_{2}\left\vert -\right\rangle _{3}+ \\
&&\sqrt{\frac{1}{3}}\left( \alpha \left\vert g\right\rangle _{1}+\beta
i\left\vert e\right\rangle _{1}\right) \left\vert \Phi \right\rangle _{23}.
\nonumber
\end{eqnarray}

\textit{Step (9)}: Let SQUID1 interact with cavity for a time $t=\pi
/2\lambda $. We have:
\begin{eqnarray}
&&\sqrt{\frac{2}{3}}\left( \alpha \left\vert 1\right\rangle _{f}+\beta
\left\vert 0\right\rangle _{f}\right) \left\vert -\right\rangle
_{2}\left\vert -\right\rangle _{3}+ \\
&&\sqrt{\frac{1}{3}}\left( \alpha \left\vert 0\right\rangle _{f}+\beta
\left\vert 1\right\rangle _{f}\right) \left\vert \Phi \right\rangle _{23}.
\nonumber
\end{eqnarray}%
\

\textit{Step (10)}: Make CNOT operations on SQUID2 and SQUID3 respectively,
then we have:
\begin{eqnarray}
&&\alpha (\sqrt{\frac{2}{3}}\left\vert +\right\rangle _{2}\left\vert
+\right\rangle _{3}\left\vert A_{\perp }\right\rangle +\sqrt{\frac{1}{3}}%
\left\vert \Phi \right\rangle _{23}\left\vert A\right\rangle )+ \\
&&\beta (\sqrt{\frac{2}{3}}\left\vert -\right\rangle _{2}\left\vert
-\right\rangle _{3}\left\vert A\right\rangle +\sqrt{\frac{1}{3}}\left\vert
\Phi \right\rangle _{23}\left\vert A_{\perp }\right\rangle ),  \nonumber
\end{eqnarray}%
where $\left\vert A_{\perp }\right\rangle =$ $\left\vert g\right\rangle
_{1}\left\vert 1\right\rangle _{f}$ and $\left\vert A\right\rangle =$ $%
\left\vert g\right\rangle _{1}\left\vert 0\right\rangle _{f}$. We notice
that Eq. (22) accords with the optimal $1\rightarrow 2$ UQCM.

\bigskip

\section{\protect\bigskip Discussion and conclusions}

In conclusion, we have proposed a scheme for realizing a $1\rightarrow 2$
universal quantum cloning machine (UQCM) with superconducting quantum
interference device (SQUID) qubits, embeded in a high-Q cavity. The SQUID
qubits can be manipulated either by the cavity field, or by classical
microwave pulses. The interaction between the $\left\vert g\right\rangle $ $%
\leftrightarrow $ $\left\vert e\right\rangle $ transition and the cavity
field can be easily "turned-on" or "turned-off" by adjusting the level
spacings of the SQUID, which can be achieved by changing the external flux $%
\Phi _{x}$ and the critical current $I_{c}$ \cite{Yang2003}. The microwave
pulses are controlled by external equipments, and can manipulate SQUID
qubits resonantly or with large detuning. In our scheme, the information is
encoded on the $\left\vert g\right\rangle $ and $\left\vert i\right\rangle $
levels, and the decay from the level $\left\vert i\right\rangle $ should be
very small, since no direct tunneling between the two levels are required%
\cite{Yang2004}. Two-photon Raman resonance processes are used to achieve
the $\left\vert g\right\rangle $ $\leftrightarrow $ $\left\vert
i\right\rangle $ transition, which has been shown faster than the
conventional gate operations \cite{Yang2004}.

Different from Rydberg atom, the SQUID used in our scheme is a kind of
"static qubit", and we have no need to trap the qubits with complex
technology. In previous proposals with Rydberg atoms, the atoms flying with
a velocity of $v=500m/s$ \cite{Milman2003} are hard to be trapped, so at
least two cavities or two field-modes are needed in the those schemes.
However, the embed SQUID can be manipulated with the cavity field for
several times, as a result, only one cavity is needed in our UQCM scheme.
Neither atom-velocity-selected device nor passing qubits detection is needed
in our scheme, and the "static" copies can easily be utilized after UQCM, so
it has advantages in experimental realization and further utilization.

\ \

\

\textbf{Acknowledgement}

\ \ \ \

This work was supported by the National Natural Science Foundation of China
under grant numbers 60578055 and 10404007.

\bigskip

Figure captions

FIG.1: The SQUID qubit driven by two
classical microwave pulses. The $\Lambda $-type three levels are denoted as $%
\left\vert g\right\rangle $, $\left\vert i\right\rangle $\ and\
$\left\vert e\right\rangle $, respectively. The $\left\vert
g\right\rangle \leftrightarrow $ $\left\vert e\right\rangle $ and
$\left\vert i\right\rangle \leftrightarrow $ $\left\vert
e\right\rangle $ transition are
coupled to classical microwave pulses 1 and 2 with identical detuning $%
\Delta $, where $\protect\omega _{1}$ and $\protect\omega _{2}$ are
the frequencies of the pulses.

\bigskip

FIG.2: Three identical SQUID qubits are embed in a high-Q cavity.
The coupling constants of SQUIDs and the cavity field are adusted to
be the same. Every SQUID can be manipulated by the cavity field or
by classical microwave pulses independently.

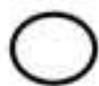
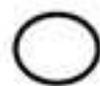
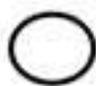
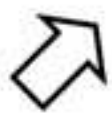
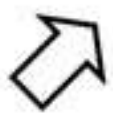
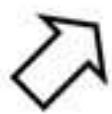
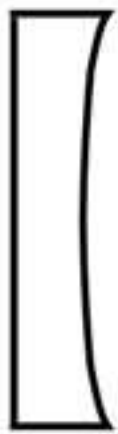
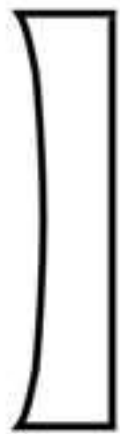
\end{document}